# Simultaneous source separation of unknown numbers of single-channel underwater acoustic signals based on deep neural networks with separator-decoder structure

Qinggang Sun[a], Kejun Wang [a,b,c*]

[a] *College of Intelligent Systems Science and Engineering, Harbin Engineering University, Harbin, 150001, China*
[b] *Key Laboratory of Intelligent Detection in Complex Environment of Aerospace, Land and Sea, Zhuhai, 519088, China*
[c] *School of Information Technology, Beijing Institute of Technology, Zhuhai, Zhuhai, 519088, China*

**Abstract**

The separation of single-channel underwater acoustic signals is a challenging problem with practical significance. Few existing studies focus on the source separation problem with unknown numbers of signals, and how to evaluate the performance of the systems is not yet clear. In this paper, a deep learning-based simultaneous separating solution with a fixed number of output channels equal to the maximum number of possible targets is proposed to address these two problems. This solution avoids the dimensional disaster caused by the permutation problem induced by the alignment of outputs to targets. Specifically, we propose a two-step learning-based separation model with a separator-decoder structure. A performance evaluation method with two quantitative metrics of the separation system for situations with mute channels in the output channels that do not contain target signals is also proposed. Experiments conducted on simulated mixtures of radiated ship noise show that the proposed solution can achieve similar separation performance to that attained with a known number of signals. The proposed separation model with separator-decoder structure achieved competitive performance as two models developed for known numbers of signals, which is highly explainable and extensible and gets the state of the art under this framework.

*Keywords:* source separation; signal separation; unknown number of sources; underwater acoustics; deep learning, machine learning

## 1. Introduction

Underwater acoustic signal separation is of great practical significance for tasks such as target recognition, target behavior analysis, and communication confrontation. Due to the influences of ocean environment noise and sea water channels, the separation of underwater acoustic signals is a challenging problem.

The blind signal separation (BSS) method is commonly used in underwater acoustic signal separation. However, it typically assumes that the number of signals is known while the signal type is unknown. It does not utilize the prior distribution of signals, and it is difficult to apply in single-channel signal separation. This paper studies the following single-channel "half-blind" signal separation problem: the maximum number of possible targets is known, the possible types of these targets are known, and the specific number of signals of each type is unknown. In this scenario, we propose a simultaneous scheme that can be applied to separate any number of signals up to the number of output channels. This is achieved by using a fixed network structure with a single-channel input and a fixed number of multi-channel outputs. This method leads to the problem that when the number of real signals is less than the number of channels, the channels that should have no signal output still have signal output. These artifacts have a great impact on the evaluation of signal separation systems. We propose a quantitative evaluation method with two metrics to deal with this problem.

---

[*] Corresponding author.
*E-mail address:* Wangkejun@hrbeu.edu.cn (K. Wang).

Deep learning has shown promising results in signal separation problems in recent years. The mainstream methods typically employ an encoder-separator-decoder structure to map the mixed signal input to the desired target signal output, e.g., dual-path recurrent neural network (DPRNN) [1], Conv-TasNet [2], and Wave-U-Net [3], along with other learning techniques. [4] and [5] proposed two two-step methods for the separation of multiple signals with known numbers. In [4], the distributions of the latent variable of each pure signal were first learned by an autoencoder or generative adversarial network, and then the optimal latent variables for reconstructing the mixtures were searched through the maximum likelihood method. In [5], the latent representations of mixtures and pure signals and the masks were first learned through an encoder-mask-decoder structure. In the second step, a separation module was inserted between the parameters of the frozen encoder and decoder to learn the mappings from mixtures to the masks of pure signals in the latent space. An approximate training method has also been used in [6] for target speaker speech extraction. In this paper, inspired by [5], we propose a novel two-step training separation model with a separator-decoder network structure. We extend these three separation models to our proposed simultaneous signal separation scheme with an unknown number of signals.

The main contributions of this study are as follows:
- We propose a solution based on a separator-decoder structural separation model with a fixed number of output channels equal to the maximum possible number of targets that are assigned to the different signal types for an unknown number of source separation. This solution does not rely on estimating the number of signals and separates all possible sources simultaneously with a fixed network structure.
- We propose a new signal separation model with a separator-decoder structure. The two parts of the model are trained separately in two steps. The proposed model has better interpretability and extendibility than competing approaches.
- We propose a new method with two quantitative metrics for evaluating the system fit with the solution.

## 2. Related works

*2.1. Underwater acoustic signal separation*

*2.1.1. Analysis based on expert knowledge*

Some studies have researched the separation of underwater signals by separating the different components of signals with different characteristics, such as spatial orientation information and category differences, in a certain signal transformation domain. Some methods separate signals directly on the feature domain based on expert knowledge [7–9]. The warping technology was used to separate dispersive time-frequency components in [7]. A depth-based method was proposed in [8], where the modified Fourier transformation of the output power of a plane-wave beamformer was used to separate the signals obtained from a vertical line array. In [9], rigid and elastic acoustic scattering components of underwater target echoes were separated in the fractional Fourier transform domain based on a target echo highlight model. However, these methods developed for specific types of signals lack generalizability. The evaluation metrics for separation performance are only compatible with the specific methods, making it challenging to compare them with other methods.

*2.1.2. Blind signal separation*

Most other algorithms rely on BSS methods [10–16]. In [10], the frequency components of the Detection of Envelope Modulation on Noise (DEMON) spectrum were used to separate signals in different directions via independent component analysis (ICA). According to the main frequency bands of different signals in a linear superposition signal, in [11], bandpass filters were used first, and then eigenvalue decomposition was employed for separation purposes [12] and [13] used the Sawada algorithm and ideal binary masking to separate artificially mixed whale songs. Polynomial matrix eigenvalue decomposition was used to deconvolute underwater signals in [14]. In [15], a method based on singular value decomposition and fast low-rank matrix approximation was proposed for ray path signal separation. A complete solution for separating and identifying hydroacoustic signals was proposed in [16] using an overdetermined BSS method based on ICA and fourth-order statistics. Most previously published studies in the field of underwater acoustic signal separation focused on the classic BSS methods, working with multiple channels.

*2.1.3. Deep learning*

Recently, a few deep learning-based approaches have been developed for underwater acoustic signal separation [17–21]. [17] proposed a network with C-RNN blocks for separating noises from two types of ships, connecting DPRNN [1] blocks and convolutional blocks from [2] in series or parallel. [18] conducted cluster analysis on features extracted by the bidirectional long short-term memory (BLSTM) network to determine the optimal binary masking value for separating ship radiated noise. [19] implemented Conv-TasNet [2] and BLSTM networks with additional Wave-U-Net [3] like encoders and decoders [22] to separate fish vocalizations from background noises. [20] utilized a BLSTM network with masks in the time domain to separate acoustic signals from ships. In [21], a BLSTM network with multi-head attention modules was used for the separation of nonlinear mixed radiated ship noise. Deep learning approaches for signal separation show promise in single-channel underwater acoustics. While these approaches have been successful when the number of signals is known, the scenario where the number of signals is unknown has not been thoroughly studied yet.

## 2.2. Signal separation with an unknown of number

Recently, some studies have also been carried out on the separation of unknown numbers of signals [12], [23–35].

*2.2.1. Estimate the number*

Some studies have solved this problem by estimating the number of signals [12], [23–30]. Among them, the number of signals was determined by independent methods in some studies, thereby simplifying the problem to a known number condition [12], [23]. In [12], a general BSS method was used after estimating the specific number of signals via the energy-based unit counting method. After independently training a counting neural network, different networks were selected for different numbers of signals in [23]. Other methods combine separation with signal number estimation [23–30]. In [24], the number of signals and a sparse matrix were estimated through clustering. [25] proposed a Bayesian model in which observation signals were reconstructed with the fewest possible frequency domain components by pruning, and the number of components was the estimated number of signals. In [26], signals were separated by spectral decomposition of the correlation matrix, and the number of signals was estimated by the eigenvalues of this correlation matrix. [27] estimated the number of signals through the rank and eigenvalues of the correlation matrix of the embedding vector and used the deep clustering method for separation. [28] combined counting and separating signals through a loss function. In [29], a counting network and a separation network shared an encoder, and different decoders were activated through the counting network to separate different numbers of mixtures. In [30], a similar technology was used for generating embedding masks. Such counting-based approaches rely heavily on the obtained counting results, and it is difficult to evaluate the separation performance of these approaches in cases with incorrect number judgments.

*2.2.2. Specially developed methods*

In addition to simplifying the problem by estimating numbers of signals, some specific approaches that are suitable for the unknown signal number situation have recently been proposed [31–35]. Some of them have solved the problem by separating signals individually [31–33]. [31] and [32] separated multiple speakers iteratively. A conditional chain model that combined a sequence model and a parallel model was proposed in [33]. These methods need to determine whether speech remains after extracting a person's speech in each iteration. [31] trained an independent binary classification network, while [32] and [33] used the energy of the remaining signal as the termination criterion. [34] proposed a constrained clustering method to assign the embedding vectors to the appropriate targets. [35] proposed a solution for separating multiple speakers and combined multiple branches for different numbers of speakers to obtain the actual output; this technique is the closest method to that in this paper. Although multiple branches can cooperate and improve the separation effect, this process also brings massive computational costs.

## 3. Methods

*3.1. Simultaneous source separation of unknown numbers without output rearrangement*

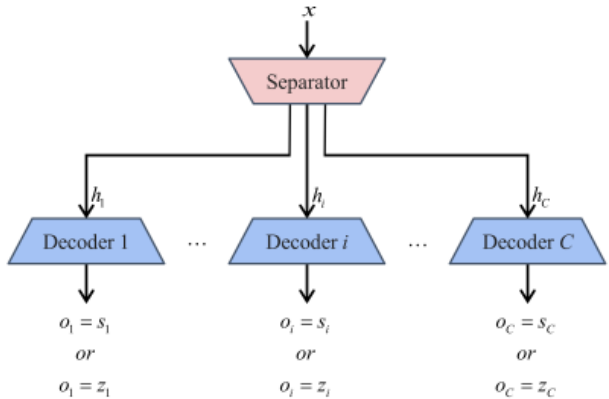

Fig. 1. (Color online) Overall framework. In the figure, $x$ represents a mixed signal, $h_i$ represents the input of the $i$–th decoder, $C$ is the total number of output channels, $o_i$ represents the desired output of the $i$–th output channel, $s_i$ represents the $i$–th type of target signal, and $z_i$ represents a sequence of zeros.

Unlike previous methods, this paper presents a simultaneous signal separation solution that can adapt to an unknown number of targets. The proposed solution does not require estimating the number of targets in the mixed signals or adjusting the network structure based on the number of target signals. As shown in Figure 1, this paper employs a network structure with a single input channel and a fixed number of output channels denoted as $C$, where $C$ represents the maximum number of targets that may be included. For any number of targets less than or equal to C, the mixed signals $x \in \mathbb{R}^T$ with a sequence length of T are consistently separated into C desired output signals $o_i \in \mathbb{R}^T, i \in \{1,...,C\}$. An approximate scheme has been recently applied to universal signal separation [36]. This fixed-number output solution presents two challenges: determining which target the output signal belongs to, especially when the actual number of targets is fewer than the number of output signals; and evaluating the performance of the system when the number of output signals does not match the actual number of targets. We will address the first problem in this subsection and propose an evaluation method for the separation system that is compatible with this solution in the next section.

Many current signal separation methods do not take into account the intended identity of the separated signals. Instead, they choose the arrangement that maximizes the evaluation metric as the separation outcome through rearrangement. This method lacks interpretability and may result in overly optimistic separation estimates. This problem is particularly noticeable when the number of target signals is unknown and the number of predicted output signals from the separation networks does not align with the actual number of targets. [37] proposed a target speaker extraction method with a known speaker identity that can avoid the permutation problem of assigning the output signals to the target speakers. In this paper, inspired by [37], each type of target signal $s_i \in \mathbb{R}^T, i \in \{1,...,C\}$ to be separated is pre-assigned to a specific channel. During the training stage, the expected output $o_i$ for each target signal is organized based on the designated channel. During the prediction stage, the output result of each channel $\hat{o}_i$ is considered as the predicted output for the specified type of target $s_i$. This approach prevents the permutation of identities of multiple output signals. In the condition of unknown numbers, for a target contained in the superimposed signal, the desired output of channel $o_i$ is the corresponding signal $s_i$; for a target that is not in the superimposed signal, the desired output of channel $o_i$ is set to a sequence of zeros $z_i \in \mathbb{R}^T, i \in \{1,...,C\}$ with the same length as that of the sample.

*3.2. Model – two-step learning approach with separator-decoder structure*

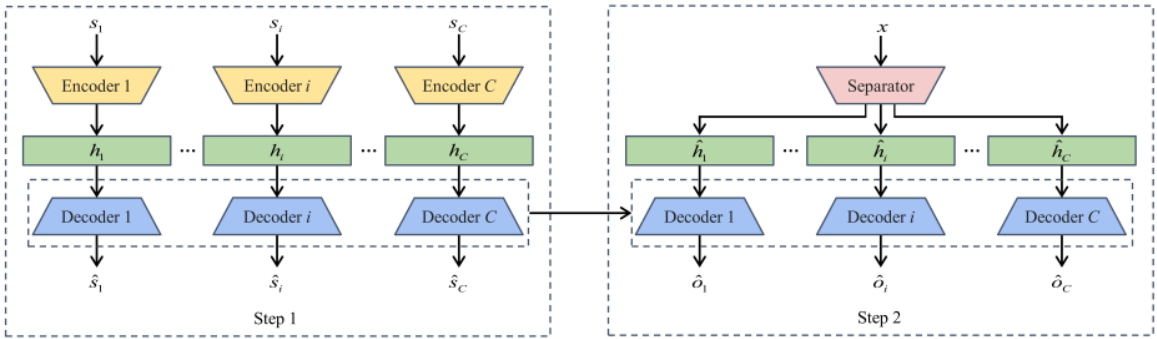

Fig. 2. (Color online) Model training process. In the figure, $s_i$ represents the i–th type of target signal, $h_i$ represents the encoded hidden variable of $s_i$, $\hat{s}_i$ represents the predicted output of the autoencoder, C is the total number of output channels, x represents a mixed signal, and $\hat{h}_i$ represents the predicted output from the separator as the learned hidden variable $h_i$. The encoder, decoder, and separator are abbreviated as *E*, *D*, and *Sep*, respectively.

As shown in Figure 2, the separation network in this paper is trained in two steps. The decoders are trained in the first step, and the separator is trained in the second step. The first step is to train the autoencoders on the training set $\{s_i \to s_i, i \in \{1,...,C\}\}$ to obtain the latent representation of each target signal $s_i \sim P(h_i), i \in \{1,...,C\}$ independently, where the encoder maps each target signal to the latent space using $h_i = E_i(s_i), i \in \{1,...,C\}$, and the decoder reconstructs each signal from the latent variables using $s_i = D_i(h_i), i \in \{1,...,C\}$. After training the autoencoders, disconnect the encoder and decoder of each autoencoder, only take out the decoder $D_i$, and fix its parameters. Then, use these decoders from the autoencoders as the decoders in the separator-decoder structured network. In the second step, a separator is trained on the training set $\{x \to o_i, i \in \{1,...,C\}\}$ to learn the mapping from the mixed signal $x$ to each hidden variable $h_i$ using $h = Sep(x), h = [h_1,...,h_C]$. If the coding vector $h$ is directly used as the training target to approximate $h$ with $Sep(x)$, the error between the output of the decoder and the target signal may be difficult to control; and the network will be difficult to train when the coding vector dimension is high. Thus, the training objective is to predict signal $\hat{o}_i$ as closely to the real $o_i$ as possible. On the test set, the network output in the second step $\hat{o}_i = D_i(\hat{h}_i) = D_i(Sep(x))$ and $o_i$ are used to evaluate the separation result.

The separator-decoder separation model with a two-step learning approach that we propose has a clearer physical meaning. While the classical model with encoder-mask-decoder structure searches for the representation of the input sample in the encoding space and the decoder parameters, our separation model finds the mapping from the input sample to the encoding space of each target signal. The autoencoder trained in the first step corresponds one-to-one with the actual type of target signal, while the training in the second step is looking for the optimal encoding that can reconstruct the target signal. The separation network is able to explicitly indicate the separated targets under an unknown number of conditions. Another benefit of training each pure signal independently is that when the task is extended to more types of signals, there is no need to train the entire system from scratch; training is only necessary for the new types and the separator.

*3.3. Loss Function*

The scale-invariant signal-to-noise ratio (SI-SNR) used in [1] and [35] and the loss function in [36] that varies with the signal type imply the condition of known numbers and types of signals on the training set. Moreover, such losses do not consider the outputs of the mute channels. Since the channels and source signal types in our solution correspond in a one-to-one manner, we use the average mean squared error (MSE) of all channels as the loss function for training:

$$Loss(o_i, \hat{o}_i; \hat{h}) = \sum_1^C \|o_i - \hat{o}_i\|_2 / C \qquad (1)$$

This technique can avoid the large computational cost required in [38] and [39].

**4. Metrics**

For the task of separating an unknown number of signals, we evaluate the quality of the separation results from two aspects. Generally, the most meaningful aspect is to evaluate the similarity between the real target signal contained in the mixture and the predicted output. We use:

$$\text{MSE}_s = \sum \|s_i - \hat{o}_i\|_2 / n_s, \text{ when } o_i = s_i \qquad (2)$$

$SDR(s_i, \hat{s}_i)$ [40] and $SI\text{-}SNR_s(s_i, \hat{s}_i)$ [36], [41]. In Eq. (2), $n_s$ represents the total number of output channels that contain the target signals from all mixed samples. On the other hand, it is necessary to evaluate how poor the obtained result is when the signal category and quantity of the predicted output are different from those of the actual signal contained in the mixed signals. Some studies have proposed counting indicators to evaluate this situation (e.g., [30], [32], [33], [36]). However, they did not give quantitative analyses of the errors of these methods. Especially in the approach of [36], if the predicted output of a silent channel is similar to other real signals, the optimal separation result obtained by permutation is not convincing. Therefore, we propose two evaluation metrics for this situation: one is the MSE between the predicted output of the mute channel and the **0** vector, as shown in Eq. (3);

$$\text{MSE}_z = \sum \|0 - \hat{o}_i\|_2 / n_z, \quad \text{when} \quad o_i = 0 \tag{3}$$

In Eq. (3), $n_z$ represents the total number of mute channels that do not contain the target signals from all mixed samples. The other is to compare the cosine similarity between the predicted output of the silent channel and the real signal of the nonsilent channel, we use the format from [36] in a different way:

$$\begin{aligned}&\text{SI-SNR}_z = 10\log_{10}[\rho^2(\hat{o}_i, s_j)/1 - \rho^2(\hat{o}_i, s_j)], \\ &\rho^2(\hat{o}_i, s_j) = (\hat{o}_i \cdot s_j)/(\|\hat{o}_i\|_2 \|s_j\|_2), \\ &i, j \in \{1, ..., C\}, \quad \text{when} \quad o_i = 0, i \neq j\end{aligned} \tag{4}$$

If the output energy of the mute channel is very small and is not similar to the real targets of other channels, we believe that the separation system can achieve good performance, with a result that is close to the real number of signals.

## 5. Experiments

To evaluate the ability of our simultaneous separation solution to cope with the signal separation of unknown numbers, we first carry out separation experiments of known numbers to give the base performances. Then, we compare the decoder-separator separation model proposed in this paper with other separation models within our framework. The superiority of our proposed evaluation method is demonstrated when evaluating the performance of the models.

### 5.1. Dataset

We select four types of signals from the *ShipsEar* dataset [42] as the sample set. Natural ambient noises and the three types of ships with the largest numbers of observations, passenger ferries, motorboats and ro-ro vessels (RORO), are represented as $s_A$, $s_B$, $s_C$ and $s_D$, respectively. The signals of $s_B$, $s_C$ and $s_D$ are superimposed on each other as $s_{BC}$, $s_{BD}$, $s_{CD}$ and $s_{BCD}$. Thus, a multitarget signal sample set containing 1–3 signals per sample is obtained. When there are multiple ship target signals in the mixed signal, we hope that the separation system can separate each target signal; when there is only one ship target signal in the mixed signal, we hope that the system can reproduce the ship signal; when there is no ship target in the mixed signal, we hope that the system can output the natural ambient noise. We maintain the original sampling rate of 52,734 Hz, and each sample is clipped with a length of 200 ms, which is 10,547 sampling points. When superimposing the signals, the SNRs of $s_B$ to $s_C$ and $s_B$ to $s_D$ are random values according to the uniform distribution ranging from -5 dB to 5 dB, so the SNRs of $s_C$ to $s_D$ follow a triangular distribution of -10 dB to 10 dB. After randomly removing and balancing the samples of eight categories, we randomly divide them into a training set, validation set and test set according to proportions of 60%, 20% and 20%, and the numbers of samples contained in these sets are 24,440, 8,144, and 8,160, respectively.

## 5.2. Comparison of Source Separation Models

### 5.2.1. Model 1 – one-step learning approach with a single encoder and multiple decoders

Similar to [35], [36] and [43], we train a network with a single-encoder and multiple-decoder structure that directly learns the mappings from an unknown number of mixed signals $x \in \mathbb{R}^T$ with a length of $T$ to the desired output signals with $C$ channels $o_i \in \mathbb{R}^T, i \in \{1,...,C\}$ simultaneously. The decoders share an encoded vector, and each has its own set of parameters for the different target signals. On the test set, the network output $\hat{o}_i$ and $o_i$ are used to evaluate the separation result.

### 5.2.2. Model 2 – two-step learning approach with autoencoders and latent variable search

Following [4], the first step only trains autoencoders for pure source signals $s_i \sim P(h_i), i \in \{1,...,C\}$. In the second step, which is executed on the test set, the decoders with fixed parameters are taken out, and the optimization problem of searching for the best input of each decoder that can reconstruct the input mixed signal is solved. The gradient descent method is then used to solve the optimization problem of searching for the best hidden variable $\hat{h}_i$. The optimization objective is to minimize the error of the mixed real superimposed signals and the addition of the predicted outputs of all channels:

$$\min\ x - \sum_i^C D_i(\hat{h}_i) \tag{5}$$

When reconstructing the mixtures, we use the MSE as the loss function:

$$Loss(x, \hat{x}; \hat{h}) = \text{MSE}_x = \|x - \hat{x}\|_2 = \left\|x - \sum_i^C D_i(\hat{h}_i)\right\|_2 \tag{6}$$

The adder for superimposing the outputs of the channels is then removed, and the output of each decoder $\hat{o}_i = D_i(\hat{h}_i)$ is the predicted output of the separation task.

## 5.3. Experimental Configurations

### 5.3.1. Network Structures

We adopt the structures of BLSTM, dual-path recurrent neural network (DPRNN) [1], Conv-TasNet [2], and Wave-U-Net [3] modules as autoencoders. In the underwater acoustic signal separation experiment with a known number of signals, we also use multiple-decoder structures without masks: BLSTM-MD, DPRNN-MD, Conv-TasNet-MD and Wave-U-Net-MD. To be fairly compared, we only use the multiple-decoder structures in Model 1 and our separator-decoder separation model (Model 3). Since Model 2 computes each sample individually in its second step, the use of a batch normalization layer in the network is not possible. Therefore, we remove the normalization layer from the networks in Model 2. The structure of the separator in the second step of Model 3 is the same as the encoder in Model 1 and Model 2, with four extra fully connected layers. The DPRNN module has a batch-related layer that cannot be used in Model 2, and its coding vector dimensionality is too high to be applied in Model 2. In Wave-U-Net-MD, we do not use skip-connection layers between the encoder and decoder, since they cannot be used in Model 2 and Model 3.

### 5.3.2. Parameter Configurations

We use the *TensorFlow* [44] and *Keras* [45] frameworks to conduct experiments[†]. In all networks, we use a convolutional layer to encode the input audio. The kernel size of this layer is 2, the stride is 1, and the number of channels is 64. In the RNN networks, the length of chunk of the input vector is 64, and the hop size is 32.

---

[†] Code and more results available at https://github.com/QinggangSUN/unknown_number_source_separation.

Thus, the input sequence is divided into 329 chunks. The number of neurons in the hidden layer is 200. Both the encoder and decoder have 1 RNN layer. In the BLSM network, the dimension of the encoding vector is 65,800 (329 * 200), and in the DPRNN network, the dimension of the encoding vector is 1,347,584 (329 * 64 * 64). The batch size is 8. In the Conv-TasNet network, the number of filters in autoencoder is 64, the number of channels in bottleneck and skip-connection paths' convolutional blocks are both 64, the number of channels in convolutional blocks is 128, the number of convolutional blocks in each repeat is 5, the number of repeats in encoder and decoder are 1 and 2 respectively. The dimension of the encoding vector is 79,200 (660 * 120). The batch size is 6. In the Wave-U-Net network, the number of downsampling and upsampling blocks is both 4, and the initial number of filters of convolutional layers in downsampling and upsampling blocks is 24 and 5, respectively. The kernel size of the convolutional layer is 15, and the dimension of the encoding vector is 79,200 (660 * 120). The batch size is 16.

In Model 2, since the *Keras* framework does not contain a variable network input layer, we use a trick to search for the optimal input vectors of the decoders. We take a constant value vector with a length of 1 and a value of 1 as the network input, connect a trainable fully connected layer with the same number of nodes as the dimension of the coding vector, and use the trained weights as the decoder input.

We employ the Adam optimizer [46] with several different initial learning rates and select the parameters with the minimum loss of each sample in Model 2, and the minimum loss on the validation set in other experiments. When training the networks, the parameters with the best performance are listed in Table 1.

| Model | Network structure | Initial learning rate | Epoch |
| --- | --- | --- | --- |
| Known number | BLSTM | 1E-3 | 100 |
| | BLSTM-MD | 1E-3 | 100 |
| | DPRNN | 1E-3 | 100 |
| | DPRNN-MD | 1E-3 | 300 |
| | Conv-TasNet | 1E-3 | 200 |
| | Conv-TasNet-MD | 1E-3 | 200 |
| | Wave-U-Net | 1E-4 | 400 |
| | Wave-U-Net-MD | 1E-3 | 800 |
| Autoencoder | BLSTM-MD | 1E-3 | 100 |
| | DPRNN-MD | 1E-3 | 200 |
| | Conv-TasNet-MD | 1E-3 | 200 |
| | Wave-U-Net-MD | 1E-4 | 1,600 |
| Unknown number Model 1 | BLSTM-MD | 1E-3 | 100 |
| | DPRNN-MD | 1E-3 | 100 |
| | Conv-TasNet-MD | 1E-3 | 200 |
| | Wave-U-Net-MD | 1E-4 | 800 |
| Unknown number Model 2 Step 2 | BLSTM | 1E-1, 1E-2, 1E-3, 1E-4 | 100 |
| | Conv-TasNet | 1, 1E-1, 1E-2, 1E-3 | 200 |
| | Wave-U-Net | 1E-1, 1E-2, 1E-3, 1E-4 | 800 |
| Unknown number Model 3 (ours) Step 2 | BLSTM-MD | 1E-3 | 100 |
| | DPRNN-MD | 1E-3 | 200 |
| | Conv-TasNet-MD | 1E-3 | 200 |
| | Wave-U-Net | 1E-4 | 800 |

Table 1. Parameters when training the networks.

## 5.4. Experiment Results

We first conduct the underwater acoustic signal separation experiment with a known number of signals. In the experiment, we take $s_{BC}$, $s_{BD}$, $s_{CD}$, and $s_{BCD}$ as inputs and $s_B$, $s_C$, and $s_D$ for each type as the outputs. Thus, a total of four separation models are individually trained and evaluated for the four types of superimposed signals. Then, in the experiments with an unknown number of signals, we mix a total of eight types of single-target and multi-target signals as input, and the three models are implemented with four channels assigned to $s_A$, $s_B$, $s_C$, and $s_D$. The results obtained on the test set are shown in Tables 2–4 and Figures 3–5. In the tables, $x$, $o$, and $s$ represent the input, channel prediction output, and real reference target signal, respectively.

The traditional metrics to evaluate the similarity between the predicted output of the channel belonging to the target signal and the real target signal are given in Table 2 and Figures 3–5. In the experiments with known numbers, the DPRNN network performed best, obtaining the highest SI-SNR$_s$ and SDR values and the lowest MSE$_s$ values. BLSM and Conv-TasNet performed similarly, slightly better than Wave-U-Net. The multi-decoder networks performed little poor than the corresponding single-decoder networks with masks. Among them, the DPRNN-MD achieved the best performance on the test set.

In the experiment of unknown number signal separation, DPRNN-MD also performed best in Model 1 and Model 3, while BLSTM-MD performed best in Model 2. The distribution of evaluation metrics among samples is shown in Figures 2–4, which are violin plots. As seen from the results in the tables and figures, our simultaneous source separation solution reproduces the input well when only one target is present. Additionally, the separation performance of Model 3 is better than that of Model 1 and Model 2 when dealing with multiple targets in the networks, except for BLSTM-MD. Model 2 performs the worst in the experiments. We believe this is caused by the fact that Model 2 does not go through the learning process on the training set when searching for hidden variables. Therefore, the error induced when reconstructing the mixed signals accumulates as the number of signals increases, and the error of each channel is difficult to individually control. Model 3 performed closely with the separators with a known number of signals, indicating the effectiveness of the proposed simultaneous separation method for an unknown number of sources.

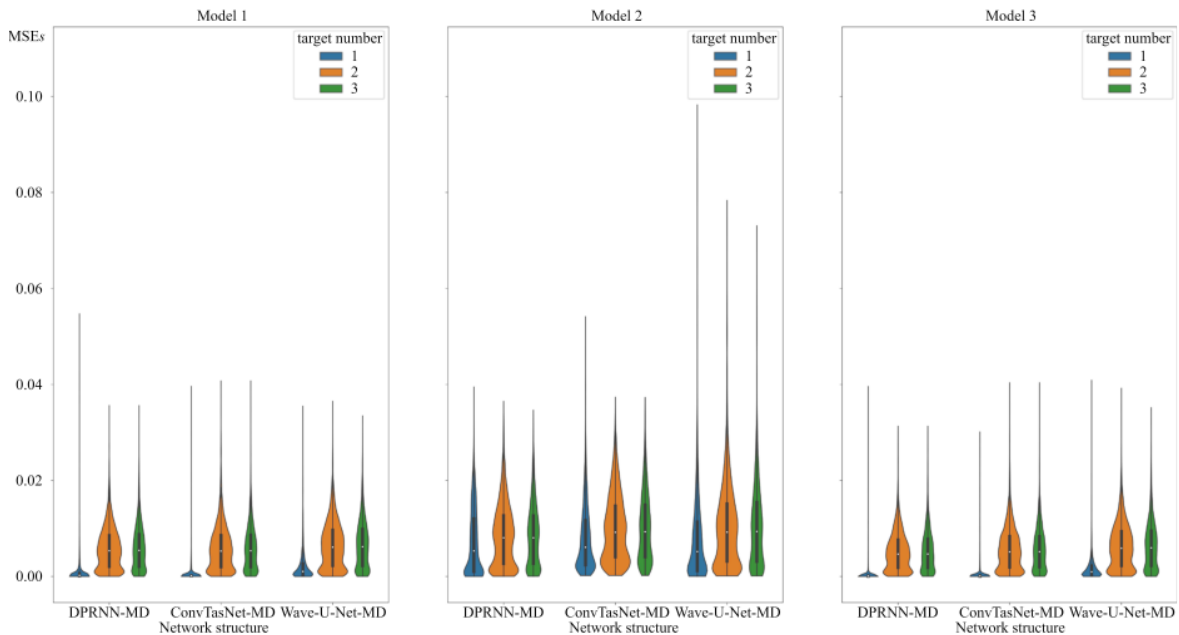

Fig. 3. (Color online) MSE$_s$ of the models for source separation with an unknown number of targets.

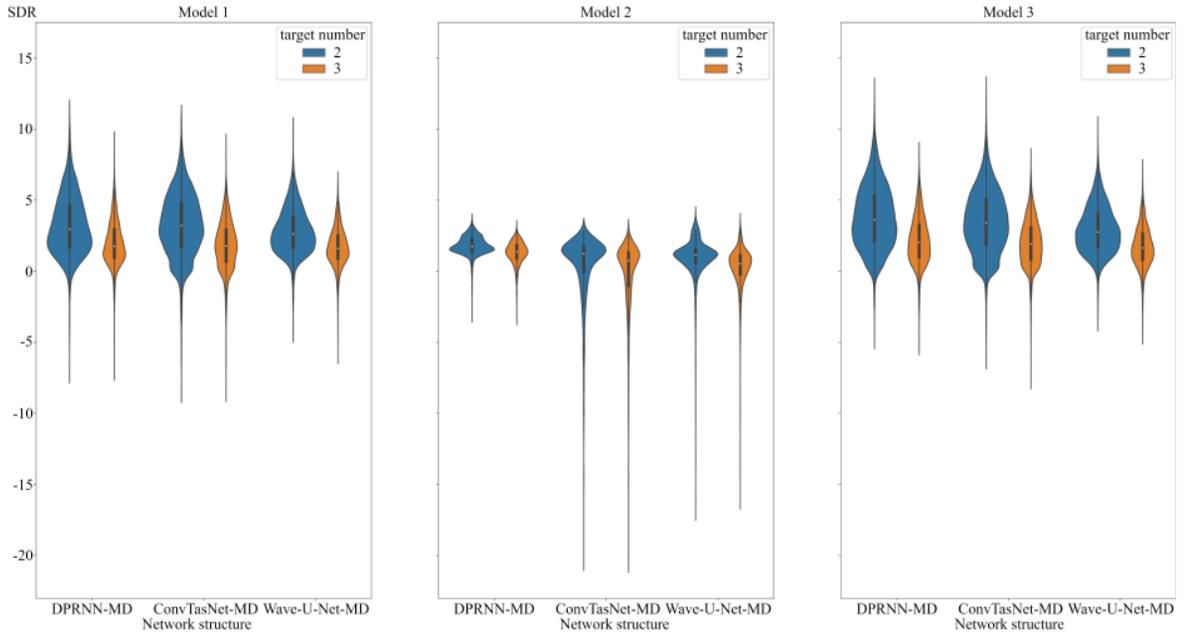

Fig. 4. (Color online) SDR of the models for source separation with an unknown number of targets.

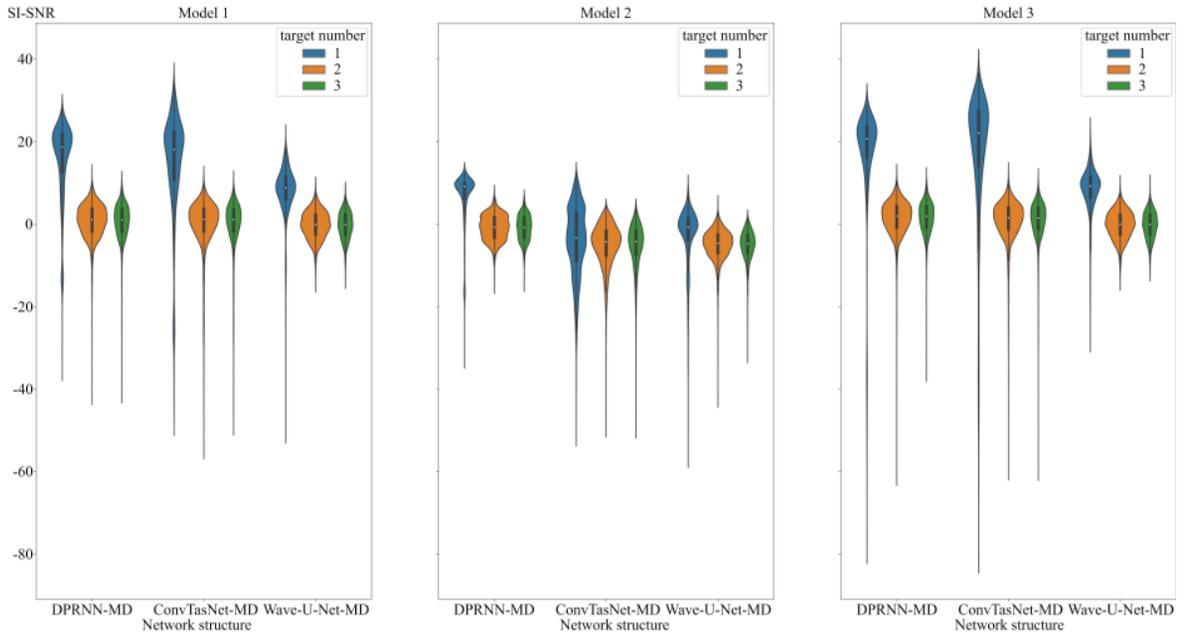

Fig. 5. (Color online) SI-SNR$_s$ of the models for source separation with an unknown number of targets.

| Model | Network structure | Target number | Metric | | |
|---|---|---|---|---|---|
| | | | $MSE_s$ | $SI\text{-}SNR_s$ (dB) | SDR (dB) |
| Known number | BLSTM | 2 | 5.27E-3 | 1.61 | 3.82 |
| | | 3 | 7.55E-3 | -1.81 | 2.14 |
| | BLSTM-MD | 2 | 5.52E-3 | 1.29 | 3.59 |
| | | 3 | 7.86E-3 | -2.10 | 1.97 |
| | DPRNN | 2 | **4.87E-3** | **2.09** | **4.18** |
| | | 3 | **7.15E-3** | **-1.36** | **2.39** |
| | DPRNN-MD | 2 | **4.85E-3** | **2.08** | **4.15** |
| | | 3 | **7.13E-3** | **-1.31** | **2.39** |
| | Conv-TasNet | 2 | 5.26E-3 | 1.64 | 3.89 |
| | | 3 | 7.61E-3 | -1.72 | 2.21 |
| | Conv-TasNet-MD | 2 | 5.30E-3 | 1.54 | 3.79 |
| | | 3 | 8.00E-3 | -2.29 | 1.92 |
| | Wave-U-Net | 2 | 5.53E-3 | 1.33 | 3.60 |
| | | 3 | 7.95E-3 | -2.09 | 1.97 |
| | Wave-U-Net-MD | 2 | 5.67E-3 | 1.15 | 3.47 |
| | | 3 | 8.02E-3 | -2.20 | 1.92 |
| Unknown number Model 1 | BLSTM-MD | 1 | 1.14E-3 | 13.30 | – |
| | | 2 | 6.03E-3 | 0.70 | 3.23 |
| | | 3 | 8.03E-3 | -2.39 | 1.88 |
| | DPRNN-MD | 1 | **8.61E-4** | **15.41** | – |
| | | 2 | **5.86E-3** | **0.68** | **3.23** |
| | | 3 | **7.79E-3** | **-2.37** | **1.96** |
| | Conv-TasNet-MD | 1 | 6.22E-4 | 14.64 | – |
| | | 2 | 5.91E-3 | -0.06 | 3.25 |
| | | 3 | 8.06E-3 | -3.31 | 1.82 |
| | Wave-U-Net-MD | 1 | 2.08E-3 | 8.05 | – |
| | | 2 | 6.57E-3 | -0.22 | 2.80 |
| | | 3 | 8.42E-3 | -3.04 | 1.70 |
| Unknown number Model 2 | BLSTM | 1 | **7.20E-3** | **6.45** | – |
| | | 2 | **8.51E-3** | **-0.89** | **1.80** |
| | | 3 | **9.47E-3** | **-3.80** | **1.30** |
| | Conv-TasNet | 1 | 8.04E-3 | -4.16 | – |
| | | 2 | 1.01E-2 | -5.09 | 0.34 |
| | | 3 | 1.17E-2 | -6.99 | -0.24 |
| | Wave-U-Net | 1 | 7.89E-3 | -2.97 | – |
| | | 2 | 1.04E-2 | -4.98 | 0.93 |
| | | 3 | 1.17E-2 | -7.09 | 0.23 |
| Unknown number Model 3 (ours) | BLSTM-MD | 1 | 1.59E-3 | 11.52 | – |
| | | 2 | 6.65E-3 | 0.13 | 2.85 |
| | | 3 | 8.57E-3 | -2.85 | 1.64 |
| | DPRNN-MD | 1 | **3.96E-4** | **16.73** | – |
| | | 2 | **5.25E-3** | **1.49** | **3.80** |
| | | 3 | **7.49E-3** | **-1.89** | **2.17** |
| | Conv-TasNet-MD | 1 | 4.70E-4 | 17.54 | – |
| | | 2 | 5.75E-3 | 0.79 | 3.47 |
| | | 3 | 7.89E-3 | -2.47 | 1.99 |
| | Wave-U-Net-MD | 1 | 1.73E-3 | 8.48 | – |
| | | 2 | 6.37E-3 | -0.21 | 2.93 |
| | | 3 | 8.38E-3 | -3.03 | 1.72 |

Table 2. Prediction outputs of the channels with targets.

The networks that performed the best among the three models based on our proposed metrics to evaluate the degree of difference between the prediction outputs of the mute channels $\hat{o}_i$ and other targets $s_j$ are given in Table 3 and Table 4. The values of $MSE_z$ and $SI\text{-}SNR_z$ of Model 1 and Model 3 are small, indicating that for target signals not included in the superimposed signal, the output energy derived from the corresponding channel is small, and the similarity with the other target signals is low. Moreover, models that perform well on traditional metrics also perform well on the new metrics. This consistency further demonstrates the rationality of the new evaluation method proposed in this paper.

| $x$ | $o$ | $s$ | SI-SNR$_z$(dB) | | | $s$ | MSE$_z$ | | |
|---|---|---|---|---|---|---|---|---|---|
| | | | Model 1 | Model 2 | Model 3 | | Model 1 | Model 2 | Model 3 |
| $s_A$ | $\hat{o}_B$ | $s_A$ | -12.67 | -1.65 | -20.17 | $s_B$ | 7.33E-5 | 4.22E-4 | 5.67E-5 |
| | $\hat{o}_C$ | | -11.81 | -0.83 | -16.02 | $s_C$ | 1.67E-4 | 4.48E-4 | 1.13E-4 |
| | $\hat{o}_D$ | | -10.36 | -1.88 | -21.37 | $s_D$ | 3.61E-5 | 4.80E-4 | 2.57E-5 |
| $s_B$ | $\hat{o}_A$ | $s_B$ | -15.05 | 7.69 | -15.35 | $s_A$ | 2.79E-5 | 6.07E-4 | 1.54E-5 |
| | $\hat{o}_C$ | | -1.64 | 7.82 | -11.94 | $s_C$ | 5.26E-4 | 6.18E-4 | 3.11E-4 |
| | $\hat{o}_D$ | | -1.84 | 8.13 | -10.83 | $s_D$ | 3.68E-4 | 7.77E-4 | 1.78E-4 |
| $s_C$ | $\hat{o}_A$ | $s_C$ | -14.05 | 7.92 | -12.66 | $s_A$ | 1.52E-4 | 7.75E-4 | 5.28E-5 |
| | $\hat{o}_B$ | | -0.78 | 8.39 | -13.74 | $s_B$ | 5.38E-4 | 7.40E-4 | 2.46E-4 |
| | $\hat{o}_D$ | | -4.11 | 7.89 | -11.30 | $s_D$ | 2.80E-4 | 8.85E-4 | 1.69E-4 |
| $s_D$ | $\hat{o}_A$ | $s_D$ | -18.29 | 8.09 | -20.29 | $s_A$ | 2.34E-6 | 7.86E-4 | 3.95E-6 |
| | $\hat{o}_B$ | | 1.17 | 9.06 | -11.11 | $s_B$ | 3.30E-4 | 8.24E-4 | 1.20E-4 |
| | $\hat{o}_C$ | | -2.94 | 8.38 | -12.21 | $s_C$ | 1.69E-4 | 8.00E-4 | 7.34E-5 |

Table 3. Prediction outputs of the mute channels with one input target.

| $x$ | $o$ | $s$ | SI-SNR$_z$(dB) | | | $s$ | MSE$_z$ | | |
|---|---|---|---|---|---|---|---|---|---|
| | | | Model 1 | Model 2 | Model 3 | | Model 1 | Model 2 | Model 3 |
| $s_{BC}$ | $\hat{o}_A$ | $s_B$ | -18.26 | -1.59 | -23.50 | $s_A$ | 1.96E-5 | 2.01E-3 | 1.13E-5 |
| | | $s_C$ | -19.17 | -0.82 | -18.17 | | | | |
| | $\hat{o}_D$ | $s_B$ | -9.12 | -0.97 | -11.10 | $s_D$ | 6.83E-4 | 2.30E-3 | 5.33E-4 |
| | | $s_C$ | -10.99 | -1.48 | -12.58 | | | | |
| $s_{BD}$ | $\hat{o}_A$ | $s_B$ | -18.51 | -0.99 | -25.39 | $s_A$ | 4.22E-6 | 1.97E-3 | 7.01E-6 |
| | | $s_D$ | -18.97 | -1.46 | -25.32 | | | | |
| | $\hat{o}_C$ | $s_B$ | -6.92 | -1.29 | -11.77 | $s_C$ | 8.35E-4 | 1.89E-3 | 5.66E-4 |
| | | $s_D$ | -9.02 | -1.05 | -14.08 | | | | |
| $s_{CD}$ | $\hat{o}_A$ | $s_C$ | -19.70 | -0.66 | -21.14 | $s_A$ | 1.08E-5 | 2.22E-3 | 8.97E-6 |
| | | $s_D$ | -18.69 | -1.88 | -25.12 | | | | |
| | $\hat{o}_B$ | $s_C$ | -5.92 | -1.35 | -11.14 | $s_B$ | 8.40E-4 | 2.08E-3 | 6.18E-4 |
| | | $s_D$ | -5.85 | -0.86 | -11.53 | | | | |
| $s_{BCD}$ | $\hat{o}_A$ | $s_B$ | -20.48 | -4.32 | -29.26 | $s_A$ | 1.21E-5 | 3.11E-3 | 1.06E-5 |
| | | $s_C$ | -20.48 | -3.54 | -24.15 | | | | |
| | | $s_D$ | -20.39 | -4.62 | -28.47 | | | | |

Table 4. Predict output of mute channels when input multiple targets.

The results show that the proposed solution for unknown number source separation can achieve similar performance to that attained in a situation with a known number of signals, and the outputs of silence channels are also good. The proposed model performs significantly better than Model 2 and slightly better than Model 1, which obtained the state of the art in this framework.

*5.5. Discussions*

In addition to performance, we have some other interesting findings. In [47], the authors proved that the decoder is more important in the encoder-mask-decoder model. In the separation model proposed in this article, the decoder parameters are trained from single target signals in the first step and are not adjusted in the second step, constituting a significant portion of the network. In the second step, only the separator in the first half of the network plays a role in source separation, which suggests that the decoder is not crucial in the separator-decoder separation model. Moreover, the autoencoder in the first step of our model is easy to train, reducing the overall learning complexity of the model. Model 2 and Model 3 are two distinct methods for searching for the best hidden variables. Model 2 is more computationally intensive as it needs to be computed on each sample. Additionally, in the first step of our model, the network only learns the pure target signal without receiving a **0** vector input. Despite this, the network exhibits robustness and can effectively reconstruct the **0** vector in the second step.

Notably, as in [32], [35], and [36], Model 1 and Model 3 need to learn the mappings from mixed signals to target signals on the training set. Model 2 only needs to learn the pure target signals on the training set and directly reconstruct the mixed signals on the test set, so it can solve the task of separating an unknown number of signals in the whole process.

Existing methods for separating an unknown number of signals have not focused on the situation where the mixed signal is superimposed from more than one target of the same type. Limited by the fact that each branch of the network corresponds to a specific signal class, the solution in this paper needs multiple identical decoders and searches for different latent variables for different targets. Therefore, modifications are necessary for the methods described in this paper. A possible solution is to determine the types of signals in them and the number of signals of each type through an independent network, which is still a difficult problem [48] for *ShipsEar* [42] dataset. Another possible solution is to enable the coexistence of multiple identical decoders and dynamically adjust the number of various types of decoders by combining the numbers, seeking the combination with the lowest loss as the optimal estimation of the target signal type. Alternatively, multiple different autoencoders can be trained independently to learn the specific features of multiple different targets of the same kind.

## 6. Conclusion

To solve the single-channel underwater acoustic signal separation problem with an unknown number of signals, we propose a simultaneous solution based on separator-decoder structure with a fixed number of output channels, which can separate any number of signals up to the number of output channels. Experiments conducted on an artificially superimposed and mixed hydroacoustic dataset from *ShipsEar* with 1–3 targets show that the proposed solution can achieve a similar separation performance to that attained under the condition with a known number of signals. We also proposed a new two-step separation model based on separator-decoder network structure, which is highly explainable and extensible, and obtained the state of the art under this framework. Moreover, we propose a new quantitative metric for evaluating the separation system fits with the solution.

**Acknowledgements**

This work was supported by the Science and Technology on Underwater Test and Control Laboratory under

Grant YS24071804.